\documentclass[twocolumn,amsmath,amssymb,prl]{revtex4}
\usepackage{color}
\usepackage{graphicx}
\usepackage{dsfont}

\newcommand{\dd}{\mathbf{d}}
\newcommand{\rr}{\mathbf{r}}

\newcommand{\rrho}{\boldsymbol{\rho}}

\newcommand{\kk}{\mathbf{k}}

\input epsf
\begin{document}
\title {Spin Waves and Dielectric Softening of Polar Molecule Condensates}
\author{Ryan M. Wilson$^{1}$, Brandon M. Peden$^{2}$, Charles W. Clark$^{1}$ and Seth T. Rittenhouse$^{2,3}$}
\affiliation{$^1$Joint Quantum Institute, National Institute of Standards and Technology and University of Maryland, Gaithersburg, MD 20899, USA}
\affiliation{$^2$Department of Physics, Western Washington University, Bellingham, WA  98225, USA}
\affiliation{$^3$ITAMP, Harvard-Smithsonian Center for Astrophysics, Cambridge, Massachusetts 02138, USA}
\date{\today}
\begin{abstract}
We consider an oblate Bose-Einstein condensate of heteronuclear polar molecules in a weak applied electric field.  This system supports a rich quasiparticle spectrum that plays a critical role in determining its bulk dielectric properties.  In particular, in sufficiently weak fields the system undergoes a polarization wave rotonization, leading to the development of textured electronic structure and a dielectric instability that is characteristic of the onset of a negative static dielectric function.
\end{abstract}
\maketitle

\emph{Introduction}--  The dielectric properties of materials have long been topics of practical interest, and there is currently a significant effort towards the production of certain metamaterials, which can exhibit novel properties including a negative dielectric function~\cite{Smith00}.  Simultaneously, breakthroughs in cold atoms physics have permitted the realization of quantum degenerate atomic gases~\cite{PhillipsRMP98,Dalfovo99,Giorgini08}, and modern efforts grant promise that heteronuclear molecules will soon be brought to quantum degeneracy~\cite{Sage05,Ni08,Aikawa10}.  Such molecules can exhibit strong electric dipolar interactions and have inherent microscopic structure that can be tuned to emulate many classes of quantum Hamiltonians and to study fundamental molecular dynamics, such as state-resolved collisions and reactions~\cite{Micheli06,Carr09,Ospelkaus10,Gorshkov11PRL,Chotia12,Quemener12,Yan13arXiv}.

In this Letter, we demonstrate that such systems can also exhibit remarkable dielectric properties.  In particular, we consider a BEC of heteronuclear diatomic molecules subject to a weak applied electric field~\cite{WilsonNJP2012,Armaitis13arXiv}, which couples low-lying states of opposite parity.  For a large class of molecules, these states are rotational in nature and are separated by an  energy $2 B_\mathrm{R}$, where $B_\mathrm{R}$ is the molecular rotational constant.  Other molecules possess a set of anomolously low energy opposite-parity electronic states, a $\Lambda$-doublet, split by an energy $2 B \ll 2 B_\mathrm{R}$ in the absence of an electric field~\cite{Browntext}.  Since the molecular polarizability $\alpha_\mathrm{mol}$ is proportional to $ B^{-1}$,  $\Lambda$-doublets can polarize in weak fields and are thus strong candidates for realizing a regime where dielectric properties result from the competition between the single and many-body interaction energies.  % For example, the hydroxyl radical (OH) has $B=884\,\mathrm{MHz}$ and $B_\mathrm{R} = 566.9\,\mathrm{GHz}$~\cite{Chemweb}.  
Recent breakthroughs in cooling techniques have allowed for the production of $\lesssim 5 \, \mathrm{mK}$ samples of OH~\cite{Stuhl12}, % $\Lambda$-doublet hydroxyl radicals (OH)~\cite{Stuhl12}, 
and with existing efforts to cool other $\Lambda$-doublets, such as ThO~\cite{Vutha11}, a new class of relevant experiments is on the horizon.

Here, we show that a BEC of diatomic molecules supports a rich spectrum of density and polarization wave quasiparticles, which play a critical role in determining the bulk dielectric properties of the system.  In particular, a novel structure emerges due to the softening of a gapped polarization wave quasiparticle in weak applied fields, which is similar to magneto-rotons in fractional quantum Hall systems~\cite{Girvin86,Pan02}.  It is also associated with the development of a negative local static dielectric function~\cite{Dolgov81,Raineri92}.  We develop a theory to describe the molecular BEC and its static dielectric properties in an oblate geometry (relevant to modern experiments~\cite{Neyenhuis12}), starting from a microscopic molecular model.

\emph{Single Molecule}--  In the absence of $\Lambda$-doubling, the low energy Hamiltonian of a heteronuclear molecule in an applied dc electric field $\boldsymbol{\mathcal{E}} = \mathcal{E}\hat{z}$ is that of a rigid rotor with a permanent electric dipole moment $d_\mathrm{mol}$ fixed to its axis~\cite{MicheliPRA07}.  In the presence of $\Lambda$-doubling, however, the relevant dipole moment is transitional between the low-lying $\Lambda$-doublet states, with a direction that is constrained to be parallel or anti-parallel with the molecular axis and a magnitude $d = c_\Lambda d_\mathrm{mol}$ where $c_\Lambda$ is a function of the electronic and rotational angular momentum quantum numbers.  We refer the reader to Ref.~\cite{Browntext}  for more in-depth discussions of $\Lambda$-doubling, but note here that in sufficiently weak fields $\mathcal{E} \ll B_\mathrm{R}/ d$, the $\Lambda$-doublet is well-described by the low-energy $SU(2)$ Hamiltonian
\begin{align}
\label{HL}
\mathrm{H}_\Lambda = B \sigma_z - d \mathcal{E} \sigma_x,
\end{align}
where $\sigma_x$ and $\sigma_z$ are Pauli matrices.  A straightforward diagonalization of~(\ref{HL}) gives the energy eigenvalues $E_\pm = \pm \sqrt{B^2+d^2 \mathcal{E}^2}$ for the eigenstates $| \pm \rangle$. 

\emph{Interactions}--   At sufficiently low energies, the interaction between heteronuclear molecules is dominantly dipolar in nature and is governed by the tensor operator~\cite{Browntext,MicheliPRA07,Gorshkov11PRA}
\begin{align}
\label{Vd1}
\mathcal{V} (\rr_1-\rr_2) = \frac{\dd_1 \cdot \dd_2 - 3 (\dd_1 \cdot \mathbf{n})(\dd_2 \cdot \mathbf{n})}{|\rr_1-\rr_2|^3}
\end{align}
where $\dd_i$ is the vector dipole operator acting on a molecule at position $\rr_i$ and $\mathbf{n} = (\rr_1-\rr_2)/|\rr_1-\rr_2|$.  % Here, we consider molecules that are tightly confined in the direction of the applied field by the potential $U(\rr) = \frac{1}{2} m \omega^2 z^2$, where $m$ is the molecular mass and $\omega$ is the trap frequency, but are free in the $xy$-plane.

For $\Lambda$-doublet molecules in the oblate geometry and applied field we consider here, Eq.~(\ref{Vd1}) is simplified as only the $z$-component of the dipole operator, $\mathrm{d}$, % has non-vanishing matrix elements. 
contributes non-negligibly.  The relevant diagonal and transitional matrix elements are $\langle \pm | \mathrm{d} | \pm \rangle= \mp d^2 \mathcal{E}  / \sqrt{d^2 \mathcal{E}^2 + B^2 }$ and $\langle \mp | \mathrm{d} | \pm \rangle = d B / \sqrt{B^2 + d^2 \mathcal{E}^2 }$, respectively.  For molecules in a finite applied field, Eq.~(\ref{Vd1}) describes interactions with both direct  and exchange character.  For the problem at hand, it will be convenient for us to work % to work in a basis where the interactions are instead purely direct.  This is accomplished by working
 in the infinite field, or ``dipole'' basis $| \kappa \rangle = R_{\kappa i} |i \rangle$, where $R$ is a rotation matrix constructed from the eigenvectors of $\sum_{ij = \pm}| i \rangle \mathrm{d}_{ij} \langle j |$.  % By construction, $\mathrm{d}$ is diagonal in this basis, and for $\Lambda$-doublet molecules it can be represented simply as $\mathrm{d} = \mathrm{d}_\Lambda \sigma_z$. 

To construct a many-body Hamiltonian, we define the spin-$1/2$ Bose field operator and expand it in the dipole basis as $\hat{\Psi}(\rr) = \sum_{\kappa =\uparrow,\downarrow} \hat{\psi}_\kappa (\rr) | \kappa \rangle$.  Including the single molecule kinetic energy term and a state-independent trapping potential $U(\rr)$, we can write the full many-body Hamiltonian as $\hat{\mathcal{H}} = \hat{\mathcal{H}}_0 + \hat{\mathcal{H}}_\mathrm{int}$, where $\hat{\mathcal{H}}_0$ is the rotated single molecule Hamiltonian,
\begin{align}
\label{H0}
\hat{\mathcal{H}}_0 = \int d\rr \hat{\Psi}^\dagger(\rr) \left[ h_0(\rr) \mathds{1} + B \sigma_x - d \mathcal{E} \sigma_z \right] \hat{\Psi}(\rr),
\end{align}
where $h_0(\rr) = -\frac{\hbar^2}{2 m}\nabla^2 + U(\rr)$, $m$ is the molecular mass and $\mathds{1}$ is the $SU(2)$ identity operator. The rotated interaction Hamiltonian can be written as
\begin{align}
\label{Hintud}
\hat{\mathcal{H}}_\mathrm{int} = \frac{1}{2} \int d\rr \int d\rr^\prime V(\rr-\rr^\prime) \hat{\Delta}(\rr) \hat{\Delta}(\rr^\prime)
\end{align}
where $\hat{\Delta}(\rr) = d \hat{\Psi}^\dagger(\rr)  \sigma_z  \hat{\Psi}(\rr)$ is the polarization density operator and $V(\rr-\rr^\prime) = (1-3 \cos^2 \theta_{\rr-\rr^\prime})/|\rr-\rr^\prime|^3$ where $\theta_{\rr-\rr^\prime}$ is the angle between $\rr-\rr^\prime$ and the $z$-axis.  In the dipole basis, the polarization density % is proportional to the $z$-component of an effective spin density, and the polarization can thus 
has the form of a longitudinal spin density, and the interactions take on a $\sigma_z^2$ Ising-like form~\cite{Goral02,Trefzger10}.

We note that the procedure for constructing a Hamiltonian with purely direct dipolar interactions is general, and also applies to rigid rotor molecules % in the appropriate trapping geometry (where only the $z$-component of the vector dipole operator is relevant) 
by choosing the two lowest lying field-dressed rotational states with zero projection of angular momentum.  However, in proceeding we consider only $\Lambda$-doublet molecules, as their zero-field splittings $B$ can be on the order of more realistic trap and interaction energies ($\sim 10 \, \mathrm{kHz}$), and therefore exhibit more interesting dielectric properties.

\emph{Bogoliubov Theory}--  To study the Bose condensed phase, we proceed along the lines of a Bogoliubov mean-field theory.  For a sufficiently tight trapping potential $U(\rr) = \frac{1}{2} m \omega^2 z^2$, where $\omega$ is the trap frequency, we limit the in-plane modes to a box of area $A$ and decompose the field operators into condensate and fluctuation terms~\cite{FWtext}, $\hat{\psi}_\kappa(\rr) = f_\kappa(z) \left(\sqrt{ n_{\kappa}}e^{i \theta_\kappa} + \frac{1}{\sqrt{A}} \sum_{\kk\neq 0} e^{i\kk \cdot \rrho} \hat{a}_{\kappa,\kk} \right)$, where $f_\kappa(z)$ is a state-dependent axial wave function, $n_{\kappa}$ is the integrated 2D condensate density and $\theta_\kappa$ is the condensate phase of state $|\kappa \rangle$.  For a single molecule, $f_\kappa(z) = \exp{\left[ -z^2 / 2 l_\kappa^2 \right]} / (\sqrt{l_\kappa}\pi^{1/4})$, where $l_\kappa = l = \sqrt{\hbar / m \omega}$ is the trap length.  In the presence of interactions, we treat $l_{\uparrow \downarrow}$ as variational parameters and integrate out the $z$-coordinate from $\hat{\mathcal{H}}$ to arrive at an effective quasi-2D (q2D) Hamiltonian.  Introducing the chemical potential $\mu$ to impose the number constraint $n_{\uparrow} + n_{\downarrow} = n_0$, we arrive at the grand canonical Hamiltonian $\hat{\mathcal{K}} = \mathcal{K}_0 + \hat{\mathcal{K}}_2 + \ldots$, where $\hat{\mathcal{K}}_2$ is quadratic in the fluctuation operators $\hat{a}_{\kappa,\kk}$.

The mean-field ground state is determined by minimizing $\mathcal{K}_0$ with respect to $l_\uparrow$, $l_\downarrow$, $n_\uparrow$ ($n_\downarrow = n_0 - n_\uparrow$) and $\theta \equiv \theta_\uparrow - \theta_\downarrow$, where
%\begin{widetext}
%\begin{align}
%\label{K0}
%&\mathcal{K}_0 = \sum_{\kappa,\lambda=\uparrow,\downarrow} \sqrt{ n_{\kappa} n_{\lambda}} \left[ \left( \frac{ \tilde{l}_{\kappa \lambda}^2 + \tilde{l}_{\kappa \lambda}^{-2}}{4} -\mu \right) \delta_{\kappa \lambda} + B \cos{\theta}  \frac{\sqrt{l_\kappa l_\lambda} }{ \tilde{l}_{\kappa \lambda}}  \sigma_{x,\kappa \lambda} - d \mathcal{E} \sigma_{z,\kappa\lambda} +\sqrt{ n_{\kappa} n_{\lambda} } \, \frac{2  \sqrt{2 \pi} d^2 }{3 \tilde{l}_{\kappa \lambda} }   (-1)^{1-\delta_{\kappa \lambda} }   \right],
%\end{align}
%\end{widetext}
\begin{widetext}
\begin{align}
\label{K0}
&\mathcal{K}_0 =  \sum_{\kappa = \uparrow, \downarrow} n_\kappa \left( \frac{ l_{\kappa}^2}{4} + \frac{1}{4 l_\kappa^2} - \mu \right)  -  d \mathcal{E} (n_\uparrow - n_\downarrow) + 2 B  \cos \theta \frac{\sqrt{n_\uparrow n_\downarrow l_\uparrow l_\downarrow}}{\tilde{l}_{\uparrow \downarrow}}  
+ \frac{4 \sqrt{2\pi} d^2}{3} \left( \frac{n_\uparrow^2}{2 l_\uparrow} + \frac{n_\downarrow^2}{2 l_\downarrow} - \frac{ n_\uparrow n_\downarrow}{\tilde{l}_{\uparrow \downarrow}} \right)
\end{align}
\end{widetext}
and $\tilde{l}_{\uparrow \downarrow} = \sqrt{(l_\uparrow^2+l_\downarrow^2)/2}$.  In Eq.~(\ref{K0}) and in the remainder of this Letter, we work in dimensionless units by scaling with the trap energy $\hbar \omega$ and length $l$.

When subject to an applied field, it is the nature of a dielectric to establish a unique field in its bulk. It is then natural to characterize a dielectric by introducing an effective polarizability, being a measure of how a constituent within the dielectric responds to an applied field.  This is in contrast to the bare polarizability, measuring the response of a constituent to a \emph{local} field, which is just the molecular polarizability $\alpha_\mathrm{mol}$ for the molecular condensate we consider here.   Defining $\beta \equiv d \mathcal{E} / B$, the effective polarizability is given by $\alpha = \frac{1}{n_0 A} \partial \langle \hat{\Delta} \rangle_0 / \partial \beta $, where $\hat{\Delta} = \int_A d\rr \hat{\Delta}(\rr)$ is the net polarization operator and $\langle \ldots \rangle_0$ denotes that we ignore fluctuations in taking this expectation value.

\begin{figure}[t]
\includegraphics[width=.85\columnwidth]{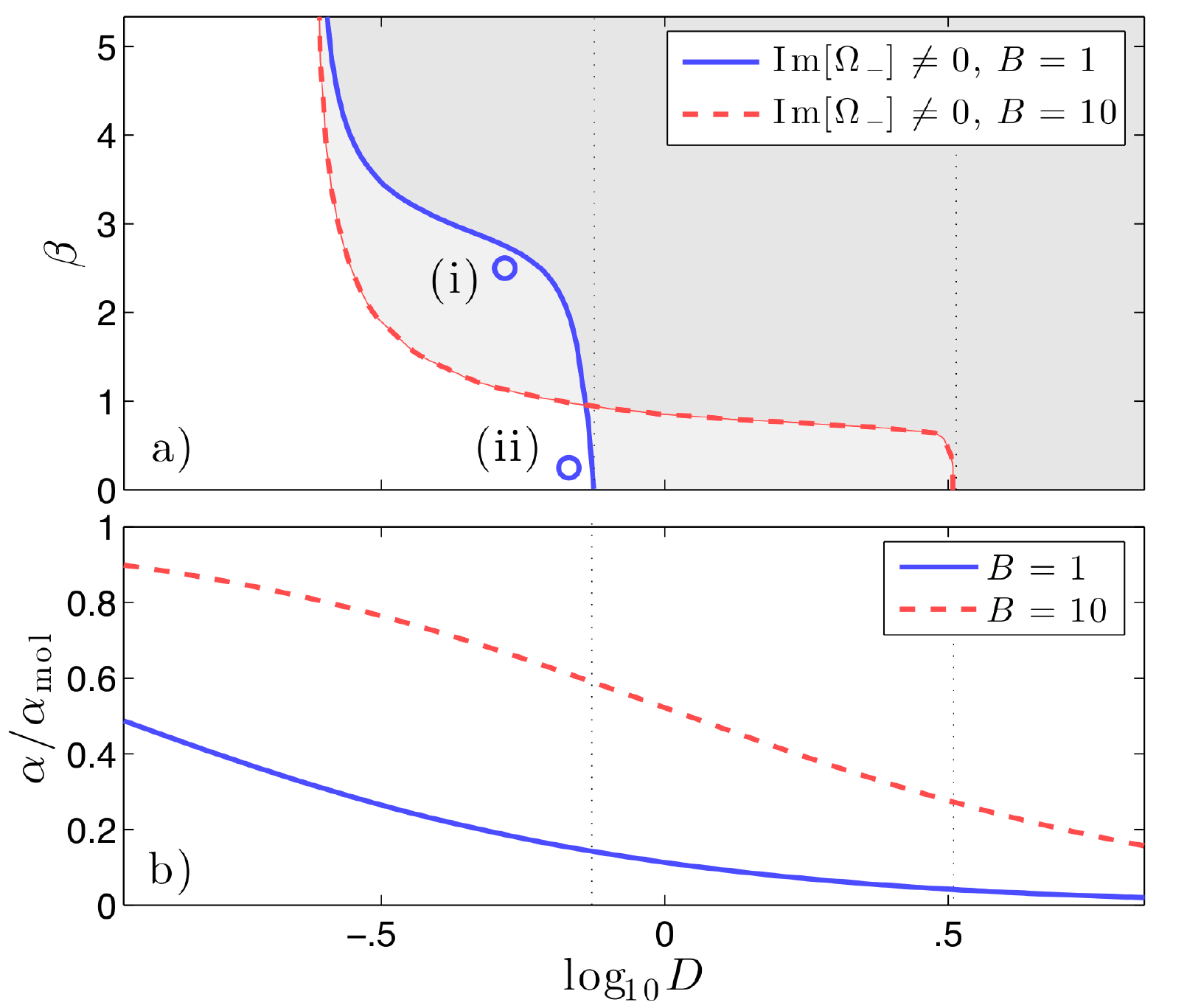} \\
\caption{\label{fig:P} (color online). a) Stability of the molecular condensate as a function of applied field ($\beta$) and dipolar interaction strength ($D$) for zero field splittings $B=1$ (blue solid line) and $B=10$ (red dashed line).  The shaded regions indicate dynamic instability.  b)  Zero-field effective polarization as a function of $D$ for $B=1$ (blue solid line) and $B=10$ (red dashed line). }
\end{figure}

Effective polarizabilities at zero field are plotted in Fig.~\ref{fig:P}(b) for $B=1,10$ as a function of the characteristic dipolar interaction strength $D \equiv n_0 m d^2 / 3 \hbar^2 l$.  For very small $D$, the polarizabilities asymptote to the single molecule value $\alpha_\mathrm{mol}$, while at intermediate $D$ the polarizabilities decrease, signifying the emergence of dielectric behavior.  At these interaction strengths, the fields produced by the dipoles themselves sum to screen the applied field in the bulk of the condensate, and the polarization is thus suppressed (in a self-consistent manner).  In the opposite limit of very large $\beta$,  the condensate fully polarizes.  This is the limit of the more familiar scalar dipolar condensate that has attracted significant interest in recent years~\cite{Santos00,Santos03,Ronen06,Wilson08,Lahaye09,Ticknor11,Boudjemaa13,Bisset13}.

In most classical dielectrics, this level of theory gives a good qualitative description of the bulk static dielectric behaviors.  However, the molecular condensate is a superfluid with ``mesoscopic'' structure in its low lying quantum fluctuations. Such fluctuations, which emerge from the condensed state as density and polarization (spin) quasiparticles, play a critical role in determining bulk behaviors.  The quasiparticles are described by the $\hat{\mathcal{K}}_2$ Hamiltonian, which we diagonalize analytically through a series of canonical transformations to arrive at the form $\hat{\mathcal{K}}_2 = \sum_{\gamma=\pm} \sum_{\kk>0} \Omega_\gamma(\kk) \hat{b}^\dagger_{\gamma,-\kk} \hat{b}_{\gamma,\kk}$.  The quasiparticle creation and annihilation operators are related to the single particle operators by the transformation $\hat{a}_{\kappa,\kk} = \sum_{\gamma = \pm }u_{\kappa,\gamma} (\kk) \hat{b}_{\gamma,\kk} + v^*_{\kappa,\gamma}(\kk) \hat{b}^\dagger_{\gamma,-\kk}$, where $u_{\kappa \gamma}$ and $v_{\kappa \gamma}$ are coherence factors representing the quasiparticle and quasihole amplitudes of a fluctuation in state $|\kappa\rangle$.  The general procedure for such a transformation is detailed in~\cite{Tommasini03}, though the momentum dependence of the dipolar interations requires further consideration.  We detail the diagonalization of $\hat{\mathcal{K}}_2$ and give explicit analytic forms for the spectra $\Omega_\pm$ and the Bogoliubov coherence factors elsewhere~\cite{Pedenfuture}.

\begin{figure}[t]
\includegraphics[width=.85\columnwidth]{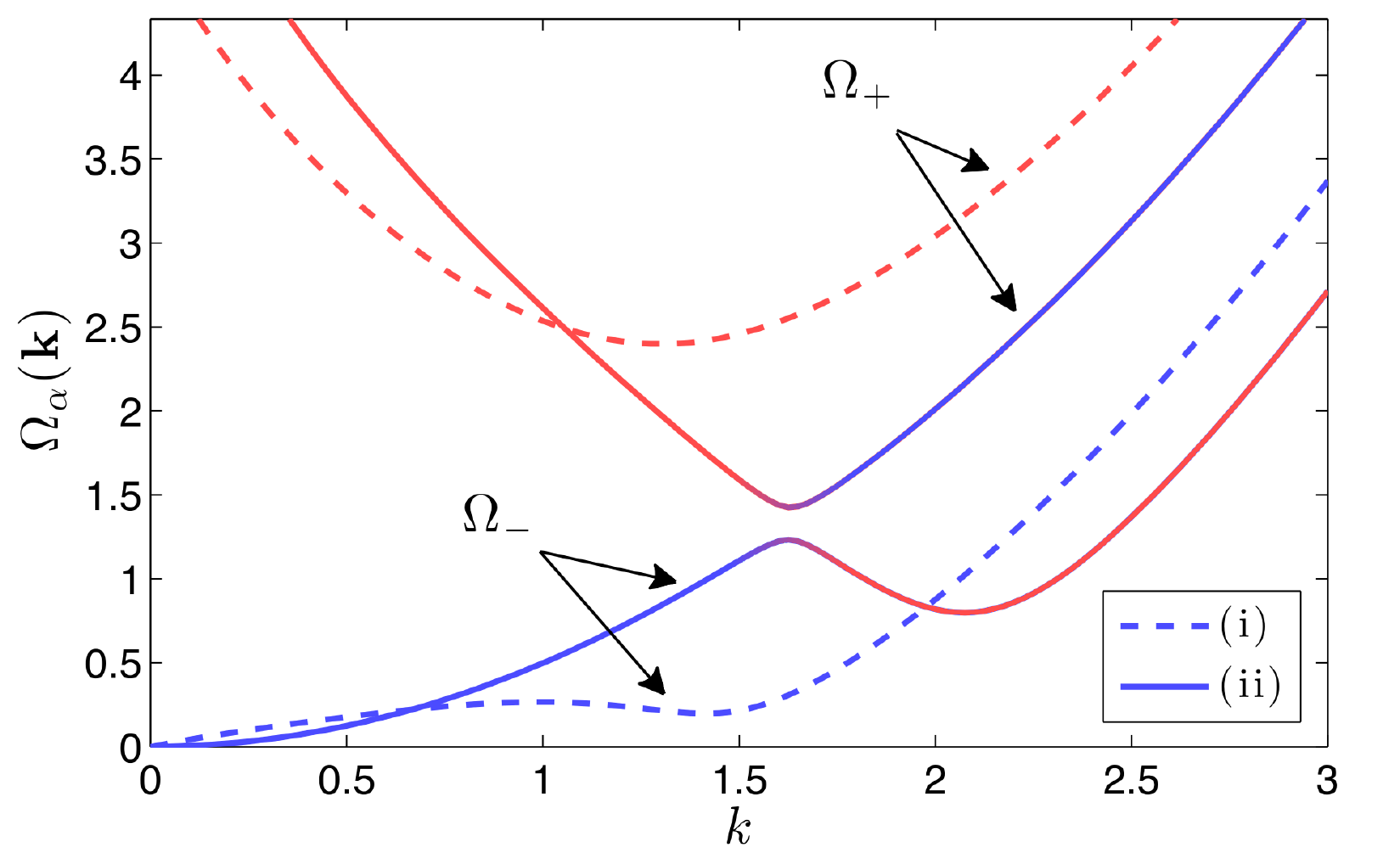}
\caption{\label{fig:disps} (color online).  Example quasiparticle dispersions $\Omega_\pm$ for the parameters (i) (dashed lines) and (ii) (solid lines), labeled in Fig.~\ref{fig:P}(a) and defined in the text.  The red coloring indicates spin-wave character and the blue coloring indicates density-wave character.  }
\end{figure}

The dispersions $\Omega_\pm$ contain a wealth of information regarding the propagation of density and polarization waves in the molecular condensate, and reveal the nature of condensate instabilities when they develop non-vanishing  imaginary values.  At small $\kk$, $\Omega_-$ describes long-wavelength density fluctuations and is gapless while $\Omega_+$ describes long-wavelength polarization (spin) fluctuations and is gapped, which we expect due to the non-commutation $[\hat{\Delta} , \hat{\mathcal{H}}] \neq 0$.   The zero-momentum gap $\Omega_+(0)$ is precisely  the energy cost of flipping a condensate molecule from the state $| \! \uparrow \, \rangle$ to the state $|\! \downarrow \, \rangle$, and this gap increases with $\beta$ until only $\Omega_-$ is energetically relevant.  In this case, $\Omega_-$ exhibits the behavior expected of a polarized scalar dipolar condensate in a q2D geometry, undergoing a rotonization for sufficiently large interaction strengths. %  In the opposite limit of small $\beta$, however, the density wave rotons are strongly suppressed by the screened interactions.

We map the stability of the q2D molecular condensate as a function of $\beta$ and $D$ in Fig.~\ref{fig:P}(a) for zero-field splittings $B=1,10$ (in units of $\hbar \omega$).  The convex instability features at larger values of $\beta$ indicate the softening of the density-wave roton.  As expected, this feature vanishes at small $\beta$, where the condensate is unpolarized.  Curiously, instead of globally stabilizing at small $\beta$, a new instability appears, which is present at all $\beta$ above a critical dipolar interaction strength $D_\Delta$.

We plot example $\Omega_\pm$ dispersions in Fig.~\ref{fig:disps} for the parameters $B=1$ and (i) $\beta=2.5$, $D=0.5$ and (ii) $\beta=0.25$, $D=0.7$.  These points are indicated by the blue circles in Fig.~\ref{fig:P}(a).  The dispersion (ii), lying just at the threshold of the small-$\beta$ instability, exhibits an avoided crossing that drives $\Omega_-$ to a local roton-like minimum at finite $k$.  The avoided crossing exchanges the density and spin wave characters of the $\Omega_\pm$ dispersions, and $\Omega_-$ takes on a spin wave character for larger $k$.  The exchange is more succinct for smaller $\beta$, where the avoided crossing becomes sharper.  The instability occurs when this spin wave roton softens, indicating that the homogeneous condensate collapses with a spin-wave pattern, into domains of field aligned and anti-aligned dipoles.  

An intuitive picture of this instability is gained by considering the relevant interaction energies at play when such domains are formed.  While their formation costs kinetic energy, adjacent domains of anti-parallel dipoles will attract, while individual domains can self-attract in the longitudinal direction if their transverse extent is small.  When $D$ is sufficiently large, these attractive interaction energies overcome the kinetic energy, and the homogeneous condensate destabilizes.

\begin{figure}[t]
\includegraphics[width=\columnwidth]{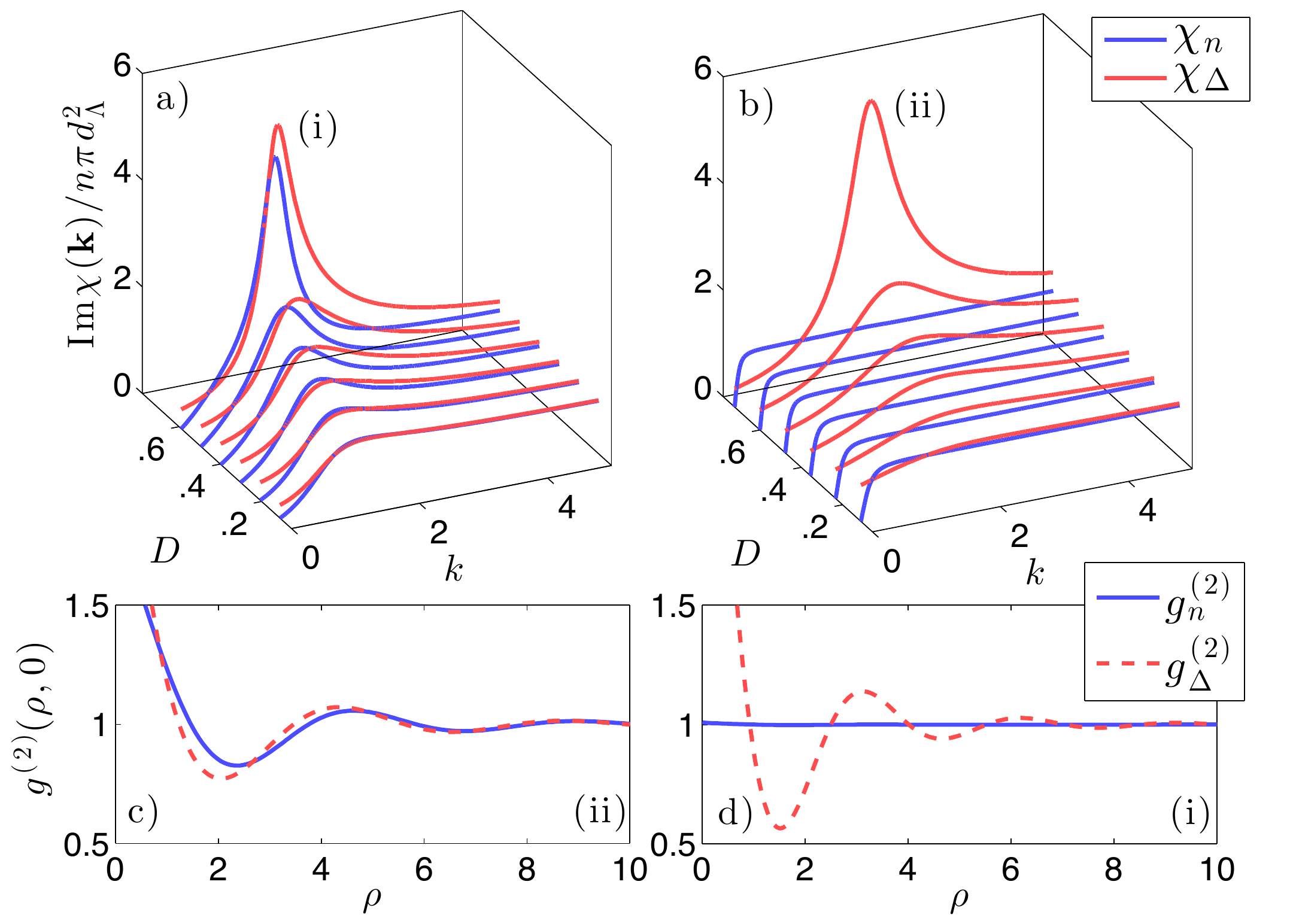}
\caption{\label{fig:sg} (color online).  Plots  a) and b) show static $k$-space density and spin (polarization) wave susceptibilities for the parameters (i) and (ii), respectively. Plots c) and d) show real-space density-density and spin-spin correlation functions for the parameters (i) and (ii), respectively.  Parameters (i) and (ii) are indicated in Fig.~\ref{fig:P} and are defined in the text. }
\end{figure}

This interpretation is clarified by an analysis of the condensate susceptibilities and two-point correlation functions.  The density wave and dielectric susceptibilities characterize the density and spin wave responses to small perturbations in the terms proportional to $\mathds{1}$ and $\sigma_z$ in the Hamiltonian~(\ref{H0}), respectively, and are given by $\tilde{\chi}_n(\kk) = i\pi \sum_{\kappa,\lambda = \uparrow,\downarrow} S_{\kappa \lambda} (\kk)$ and $\tilde{\chi}_\Delta(\kk) =  i\pi d^2 \sum_{\kappa,\lambda = \uparrow,\downarrow}  (-1)^{1-\delta_{\kappa \lambda}} S_{\kappa \lambda} (\kk)$, where $S_{\kappa \lambda}(\kk)$ are the static structure factors of the condensate in the $\theta = \pi$ ground state~\cite{Dalfovo92,Zambelli00}, 
\begin{align}
S_{\kappa \lambda}(\kk) = \sum_{\gamma=\pm} (-1)^{1-\delta_{\kappa \lambda}} \delta n_{\kappa \gamma}(\kk) \delta n^*_{\lambda \gamma}(\kk),
\end{align}
and $\delta n_{\kappa \gamma}(\kk) = \sqrt{n_\kappa} (u_{\kappa \gamma}(\kk) + v_{\kappa \gamma}(\kk))$.  We plot these susceptibilities in Figs.~\ref{fig:sg}(a) and~\ref{fig:sg}(b) for the parameter sets (i) and (ii).  For the parameters (i), $\Omega_+$ is sufficiently gapped at all relevant values of $k$ (see Fig.~\ref{fig:disps}) that the susceptibilities are density wave dominated.  Also, the condensate is polarized with $n_\uparrow \sim 3 n_\downarrow$ at these parameters, so fluctuations in the condensate density carry trivial longitudinal spin fluctuations.  We thus see peaks develop  in both the $\chi_n$ and $\chi_\Delta$ susceptibilities as $D$ is increased from $D=0.1$ to $0.6$ in Fig.~\ref{fig:sg}(a), where the back-most susceptibilities are plotted for the parameter set (i).  By contrast, for the parameters (ii), the density wave susceptibilities $\chi_n$ are nearly featureless for all $D$, while the spin wave susceptibilities develop a strong peak at finite $k$ with increasing $D$.  The spin wave instability corresponds to a divergence in this susceptibility at the critical interaction strength $D_\Delta$.  In this sense, the instability is characterized by the onset of spin structure without an accompanying mechanical structure, signifying a novel dielectric response of the molecular condensate.  We thus term this instability the dielectric instability.

Further insight into the nature of the dielectric instability is gained by inspection of the spin-spin correlation function, which up to a short range part $\propto \delta(\rho)$ is given by  $g^{(2)}_\Delta(\rho) = 1+ \frac{1}{n_0} \int \frac{d\kk}{(2\pi)^2} e^{-i\kk\cdot \rrho} S_\Delta(\kk) $, and we define the density-density correlation function analogously.  We plot these correlation functions in Figs.~\ref{fig:sg}(c) and~\ref{fig:sg}(d) for parameters (i) and (ii).  For parameters (ii), near the dielectric instability, the density correlations are nearly featureless, while the spin correlations exhibit strong oscillations.  The first minimum near $\rho \sim 1.5$  continues to lower with increasing $D$ until the stability threshold is reached, at which point the correlations indicate clear antiferroelectric order.  Thus, in a weak but finite applied field, the system develops a negative local static dielectric function, where locally the polarization of the condensate aligns antiparallel with the field.  

\emph{Discussion}-- In the $\beta=0$ limit, we find that the critical dipolar interaction strength $D_\Delta$ scales approximately linearly with $B$, and the critical 3D density for the dielectric instability can be estimated by $n^{(3D)}_\Delta  \sim B / 2 d^2$, where $n^{(3D)} = n/ \sqrt{\pi} l$ is the maximum condensate density achieved at $z=0$ (in the $\beta=0$ limit).  % For a trap with frequency $\omega = 2\pi \times 20 \, \mathrm{kHz}$, this gives large critical densities $n^{(3D)}_\Delta \gtrsim 10^{17} \, \mathrm{cm}^{-3}$ for the heteronuclear alkali molecules and OH, with zero-field splittings on the order of a GHz~\cite{QuemenerJCP12}.  
For a trap with frequency $\omega = 2\pi \times 20 \, \mathrm{kHz}$, this gives a critical density of just $n^{(3D)}_\Delta \simeq 1.32 \times 10^{13} \, \mathrm{cm}^{-3}$ for ThO, due to its very small $\Lambda$-doublet splitting $B \simeq 15.1 \, \mathrm{kHz}$ and large dipole moment $ d = 1.95 \, \mathrm{Db}$ ($d_\mathrm{mol} = 3.89$ and $c_\Lambda = 1/2$ for ThO in its metastable $H\, ^3\! \Delta_1$ $\Lambda$-doublet state)~\cite{Buchachenko10,Vutha11}.  Thus, $\mathrm{ThO}$, and other molecules with anomolously small splittings are ideal candidates for accessing the dielectric instability.

While here we consider a homogeneous q2D geometry, modern experiments achieve largely oblate but 3D trapping geometries using, for example, retro-reflected lasers for optical trapping in 1D lattice potentials.  Our results will hold for such trapping geometries as long as the ratio of the axial and radial trap frequencies is sufficiently large, $\omega_z / \omega_\rho \gg 1$.  In this case, the onset of the dielectric instability will occur where the density is greatest, in the center of the trap.  Also, it is notable that the peak in the dielectric susceptibility is quite apparent for moderate interaction strengths, well below the dielectric instability threshold.  Such features should be observable via optical Bragg scattering~\cite{Stenter99}.  Finally, we note that while we consider pure dipolar interactions, the low energy interactions between two polar molecules may also involve a short-range part.  In investigating the role of such interactions, we find that they shift the stability/phase diagram in Fig.~\ref{fig:P}(a), but introduce no qualitative changes.

\emph{Conclusion}-- In summary, we considered the fundamental properties of a BEC of heteronuclear polar molecules in the experimentally relevant q2D geometry, and in weak applied fields.  In the parameter regime where the trap energy, the (screened) interaction energy and the zero-field splitting of the low-lying opposite-parity molecular states are all comparable, the condensate exhibits novel dielectric behavior that emerges due to strong quasiparticle fluctuations of the condensate polarization.  In particular, the homogeneous condensate destabilizes due to a diverging dielectric susceptibility into domains of field aligned and anti-aligned dipoles, signifying the onset of a negative local static dielectric function.  This system thus represents a new class of materials exhibiting both novel dielectric and quantum mechanical behaviors.  

\emph{Acknowledgments}--  We acknowledge useful conversations with B. L. Johnson, H. R. Sadeghpour, C. Ticknor, E. Timmermans, and J. L. Bohn.  RMW acknowledges support from an NRC postdoctoral fellowship.

\end{document}